\documentclass[12pt]{iopart}
\usepackage{graphicx}

\begin{document}
\title{Critical and Gaussian conductivity fluctuations in BaFe$_{1.9}$Ni$_{0.1}$As$_2$ superconductor.}
\author{S. Salem-Sugui, Jr.$^1$, A.D. Alvarenga$^2$, R. I. Rey$^3$, J. Mosqueira$^3$, H-Q Luo$^4$, X-Y Lu$^4$}
\address{$^1$Instituto de Fisica, Universidade Federal do Rio de Janeiro,
21941-972, Rio de Janeiro, RJ, Brazil}
\address{$^2$Instituto Nacional de Metrologia Qualidade e Tecnologia, 25250-020 Duque de Caxias, RJ, Brazil}
\address{$^3$LBTS,Universidade de Santiago de Compostela, E-15782, Spain}
\address{$^4$Beijing National Laboratory for Condensed Matter Physics, Institute of Physics, Chinese Academy of Sciences, Beijing, 100190, P. R. China}
\begin{abstract}
We study fluctuation conductivity in a single crystal of BaFe$_{1.9}$Ni$_{0.1}$As$_2$ superconductor ($T_c$ = 20 K) as a function of temperature and applied magnetic field. Magneto-conductivity curves, $\Delta \sigma$ vs. $T$,  were analyzed in terms of -1/(dln($\Delta \sigma$)/d$T$) vs. $T$ plots, which allow to study different fluctuations regimes and to estimate exponent values and temperature widths of each regime. The analysis of magneto-conductivity curves evidence the existence of only two fluctuations regimes, a possible critical one (of glass-like type) going from the irreversible temperature to above $T_c(H)$ followed by Aslamazov-Larkin fluctuations in the Gaussian regime.
\end{abstract}

\maketitle 

 \section{Introduction}
The study of superconducting fluctuations above the superconducting transition temperature, $T_c$,  in iron-based superconductors \cite{1} has gain increasing attention \cite{rey1,2009,nos,jesus,phyC,magn,benfato,tesab,sust,liu1,liu2,liu3,ssTcom,pandya,pandya2,ltp,prando,f,rey2}, as these effects extend up to relatively large temperatures above $T_c$, when compared with low-$T_c$ superconductors.\cite{tinkham} These enhanced fluctuations as observed in pnictides are due to their layered structure in conjunction with their relatively large $T_c$, large values of the Ginzburg-Landau (G-L) parameter, $\kappa$, and large values of the upper critical field, $H_{c2}(0)$ (low values of the coherence length).  Studies of superconducting fluctuations, near and above $T_c$, allow to compare experiments with available theoretical predictions for critical and Gaussian fluctuations in superconductors. As fluctuations are sensitive to the pairing symmetry, it is of particular interest to study pnictides, for which multi-band (inter-band) levels participate in the pairing \cite{tesa2,review} producing a $s\pm$ symmetry (the symmetry is still not clear, and seems to depend on the doping \cite{benfato}). For instance, a  recent theoretical work  taking into account the characteristic inter-band pairing  of the pnictides and a $s\pm$  symmetry \cite{benfato} predicted that conductivity fluctuations should exhibit only one critical regime, while two critical regimes are expected near $T_c$ for a s-wave pairing (a statical one, followed by a dynamical).\cite{lobb} Superconducting fluctuations can also provide information on the system dimensionality. Analysis of critical and Gaussian fluctuations in the studied pnictides in the literature appears to support a three dimensional, 3D, behavior \cite{nos,jesus,phyC,magn,sust,ssTcom} but two dimensional behavior was observed for higher fields (above 8 T) in SmFeAs system \cite{2009} and in LiFeAs \cite{f}, and a recent theory of critical fluctuations was specially developed for pnictides considering an intermediated dimensionality between two and three for high applied magnetic fields which seems to explain high-field experimental data as well.\cite{tesab}

The Ginzburg criterium which defines the temperature width of critical fluctuations close to $T_c$ can be estimated by the relation $\vert T-T_c \vert$ $<$ 1.07x$10^{-9}(\kappa ^4T_c^3/H_{c2}(0))$ K \cite{lobb} where $H_{c2}(0)$ is the field obtained by extrapolating the linear region near $T_c$ in the $H$ vs. $T$ diagram to $T$ = 0 K. Values for  conventional superconductors are \cite{lobb} $\vert T-T_c \vert$ $<$ $10^{-6}$ K,  $\vert T-T_c \vert$ $\leq$ 1 K for high-$T_c$ cuprates \cite{klemm,paulo,AD}, but of the order of $10^{-3}$ K for pnictides.\cite{magn} This predicted narrow critical region, difficult  to be resolved experimentally, can increase orders of magnitude under the effect of a high-magnetic field that constricts the pairs to the lowest Landau level, lowering the system dimensionality.\cite{rosenstein} The latter explains the considerably large critical fluctuation regime that has been observed in pnictides under high applied magnetic fields.\cite{2009,nos,jesus,phyC,magn,sust,liu1,liu2,liu3,ssTcom} In most of the cases the study of high-field critical fluctuations is possible by checking if  experimental curves follow specific scaling laws derived from the G-L theory developed for layered systems within the lowest-Landau-level approximation, LLL, and taking in to account interactions in the fluctuation regime.\cite{ullah,rosenstein2} These scalings laws when applied to experimental data may allow to estimate many values of G-L intrinsic parameters \cite{rosenstein}, but unfortunately, do not provide even a qualitative information regarding the temperature width where these  fluctuations are important.

Conductivity fluctuations above the surperconducting transition temperature, also known as excess conductivity or paraconductivity, have been widely theoretically and experimentally studied in most of known superconductors \cite{tinkham,rosenstein}, including earlier works studying Gaussian fluctuations at zero magnetic field within the well known Aslamazov-Larkin, A-L, theory.\cite{AL,tinkham} In more recent experimental works performed in the presence of magnetic fields, fluctuation effects are interpreted in terms of extensions of the A-L theory \cite{rey1,caballeira} when in the Gaussian regime, and, near the transition temperature, by scaling-laws derived from G-L theories which include correlations in the fluctuation regime, then critical.\cite{ullah,rosenstein} For pnictides, magneto-conductivity studies were performed in SmFeAsO \cite{2009,ltp}, fluor doped SmFeAsO \cite{liu1}, (Nd,Pr)FeAsO \cite{liu2}, (Nd,Pr,Sm)FeAsO \cite{liu3},  BaFe$_{2-x}$Co$_x$As$_2$ \cite{magn}, Ba$_{1-x}$K$_x$Fe$_2$As$_2$ \cite{sust}, LaOFeAs \cite{ssTcom}, LiFeAs \cite{f} and in BaFe$_{2-x}$Ni$_{x}$As$_2$ \cite{rey1}. For most of these studies \cite{2009,liu1,liu2,liu3,magn} data were analyzed in terms of scaling-law approaches for critical fluctuations,  while data in Refs. \cite{sust}, \cite{ssTcom} and \cite{rey1}  were analysed in terms of a 3D A-L theory \cite{AL} or its extensions.\cite{rey1}

In this work we study isofield resistivity curves previously obtained \cite{rey1} in a high-quality BaFe$_{1.9}$Ni$_{0.1}$As$_2$ single crystal \cite{luo} with $T_c$ = 20.0 K and $\Delta$$T_c$ $\approx$ 0.3 K (electron-doped 122 system). We focus our analysis on the width of the critical regime above the superconducting transition, the associated critical exponents, and the evolution of this regime with the applied magnetic field. As will be shown, the results resolved for the first time the width of the critical region under applied magnetic field and its associated critical exponent.  

 \section{experimental}  
As detailed in Ref. \cite{rey1}, precise isofield resistivity, $\rho$, measurements were obtained as a function of temperature, for magnetic fields running from zero to 9 T applied parallel to the c-axis of the sample. The excess conductivity, $\Delta \sigma$, above the superconducting temperature transition is obtained from each isofield resistivity curve by subtracting the normal state resistivity, $\rho$$_n$: $\Delta \sigma$=1/$\rho$-1/$\rho$$_n$. The normal state resistivity is obtained by fitting the form $\rho$$_n$=b+c$T$+d$T$$^2$ in a selected temperature region (within a temperature width of 30 K) in the normal state. 

Figure 1 displays the resulting conductivity curves. The upper inset of Fig. 1 shows a selected resistivity curve and the corresponding fitting performed in the normal state region extrapolated to lower temperatures. The lower inset shows d$\rho$/d$T$ as obtained from the resistivity curve with $H$=0, evidencing the sharp transition. The solid line crossing the isofield conductivity curves in Fig.1 represents a virtual line separating critical and Gaussian fluctuation regimes which discussion is presented  below. Arrows in each curve of  Fig.1 indicate values of $T_c(H)$ obtained from the LLL scaling, which is also discussed below.
\begin{figure}[t]
\includegraphics[width=\linewidth]{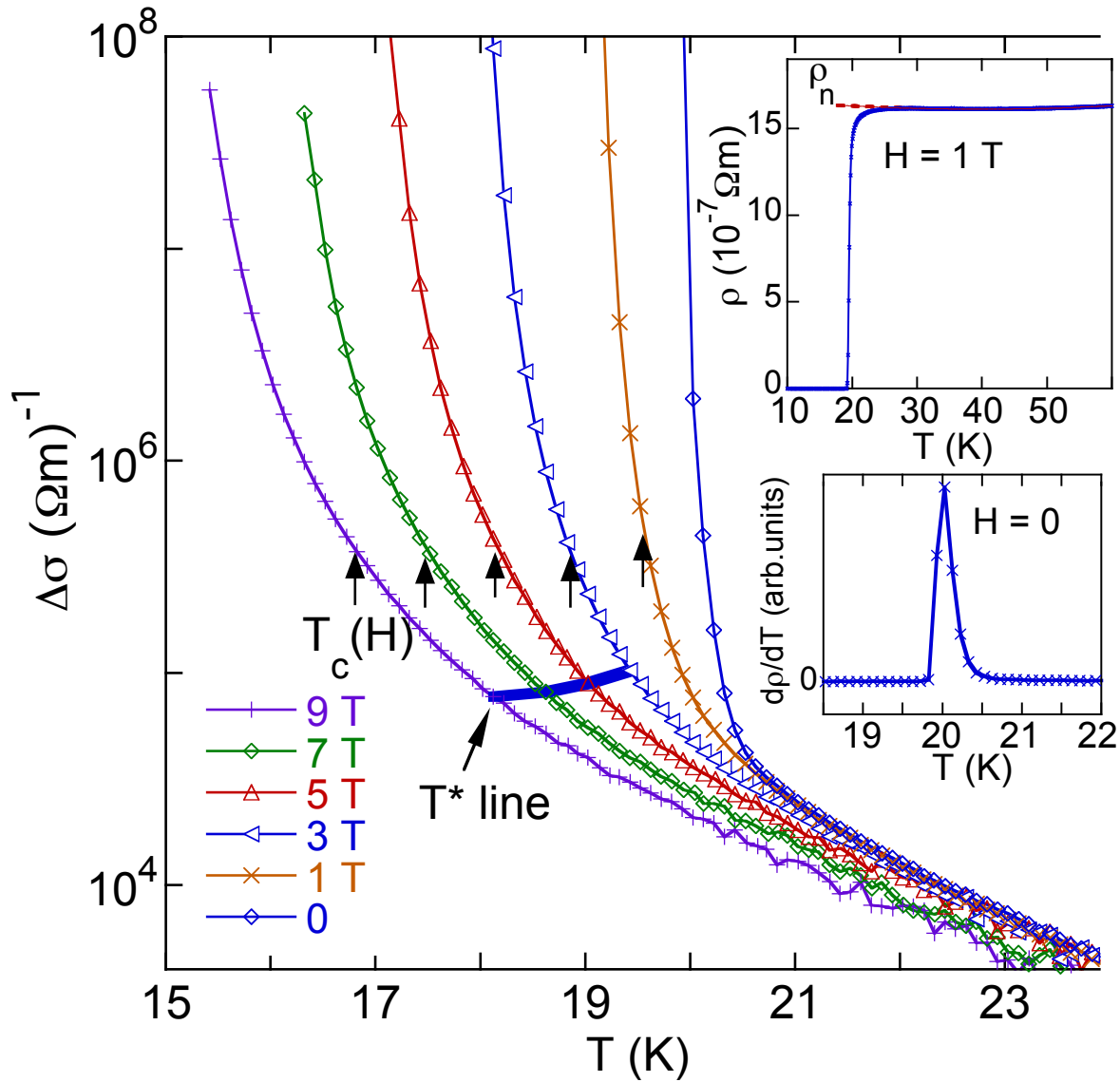}
\caption{$\Delta \sigma$($H$) vs. T curves. The upper inset shows a selected resistivity curve and the fitting performed on the normal state region,  $\rho$$_n$, extrapolated to lower temperatures. The lower inset shows d$\rho$/d$T$ for $H$ = 0.}
 \label{fig1}
\end{figure}
 \section{Results and discussion}  
From the Aslamazov-Larkin theory \cite{AL}, the excess conductivity, $\Delta \sigma$, above $T_c$ for a three dimensional superconductor is proportional to the coherence length $\xi (T)$.  Then, the excess conductivity above $T_c$ is expected to behave as \cite{lobb}
$$\Delta \sigma \equiv A(T-T_c)^{- a}$$
where A is a constant and $a$ an exponent which assumes different values depending on the regime of fluctuations. This equation is expected to hold in the presence of a magnetic field by just replacing $T_c$ by $T_c(H)$. In the absence of magnetic fields, the expected values of  the exponent $a$ are: $a$=1/2 in the mean field regime, crossing over to $a$$\approx$0.67 as temperature approaches $T_c$ corresponding to a static critical exponent, crossing over to $a$$\approx$0.33 even closer to $T_c$ corresponding to a dynamical critical exponent \cite{lobb}. The equation above can also be written as  -1/(dln($\Delta \sigma$)/d$T$)=(1/$a$)($T$-$T_c$), and analysis of experimental conductivity curves in terms of plots of  -1/(dln$\Delta \sigma$/d$T$) vs. $T$ may allow to estimate exponent values and temperature widths in the fluctuation regime.\cite{paulo} Here,  we apply this approach, among others, to analyze the conductivity curves presented in Fig.1.
 
Figure 2a and 2b show  plots of the quantity  -1/(dln($\Delta \sigma$)/d$T$) vs. $T$ as obtained from the conductivity curve for $H$ = 0 and $H$ = 1 T respectively. The derivatives dln($\Delta \sigma$)/d$T$ were calculated by using the commercial program Origin 8. As shown in Fig. 2a, it is possible to identify two distinct linear regions above $T$$\approx$19.9 K. The first region with a width $\Delta$$T$$\approx$0.6 K has an exponent $a$$\approx$0.42, which is close to the mean field value $a$=0.5 and it appears to be associated to three dimensional A-L fluctuations, as discussed above. An extrapolation of this region to the $T$ axis, sugests the value of $T_c$ = 20.2 K. The region below 20.2 K in Fig. 2a is likely to be associated with the transition width $\approx$0.3K. As expected, the experiment can not resolve any critical region near $T_c$, which would occur within a 10$^{-3}$ K width. A second region with $a$$\approx$3  starts at $T$$\approx$ 21.1 K and seems to extend up to higher temperatures although a large spread of the data is observed above 23-24 K (not shown). The inset of Fig. 2a shows the original ln$\Delta \sigma$ vs. $T$ curve of the main figure and the dashed line is the fitting of the data following the 3D A-L theory presented in Ref.\cite{rey1}. Arrows in this inset give a rough indication of the exponents $a$ in the curve. Between this two regions, the plot in Fig. 2a shows a small region with decreasing curvature from which, one can barely resolve an exponent $a$$\approx$0.85. Probably this small region has no associated exponent and just  represents a decay in the amplitude of A-L fluctuations, which is physically expected to occur as temperature increases above $T_c$  (entering in the region with $a$$\approx$3). Such decay in the amplitude fluctuations are related to the well-known short-wave length effects occurring well above $T_c$ \cite{johnson,vidal}, and explains the rather large value of the exponent $a$$\approx$3 associated with A-L Gaussian fluctuations in the corresponding region.\cite{rey1}
 
Figure 2b shows the same plot of Fig. 2a, but with a magnetic field $H$ = 1 T, which exhibits a totally different scenario. It is possible to resolve three different linear regions in this figure.  The first region, with $a$$\approx$2.4 starts just above the irreversible temperature, $T_{irr}$ ($T_{irr}$ is defined here as the temperature below which a critical current $J_c$$>$0 is established),  and ends close to the mean field transition temperature, $T_c(H)$, which is estimated by extrapolating the second regime with $a$$\approx$0.88 to the $T$-axis (marked with an arrow). Above $T_c(H)$ there are two distinct regions, the first, with $a$$\approx$0.88, is most likely to be related to the mean field A-L fluctuations under a magnetic field (similar to the A-L fluctuations with $a$$\approx$0.42 observed in Fig. 2a with $H$ = 0), and the second one, with $a$$\approx$1.4, is likely to be associated to A-L fluctuations with short-wave length effects since data in this region lies in the same region where the amplitude fluctuations decay to zero as temperature increases.\cite{rey1}

\begin{figure}[t]
\includegraphics[width=\linewidth]{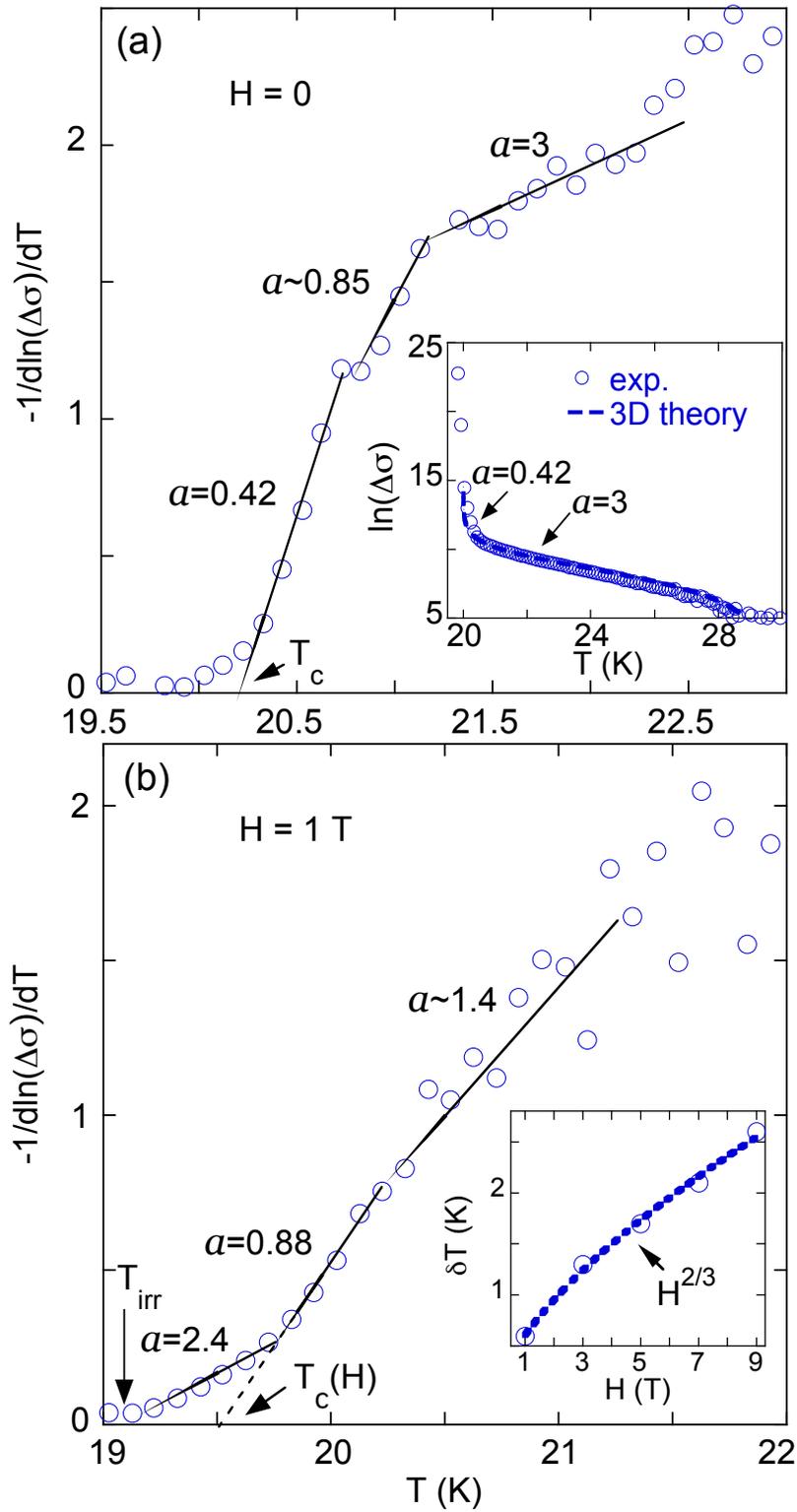}
\caption{-1/dln($\Delta \sigma$)/d$T$ is plotted against $T$ for: a) $H$ = 0; and b) $H$ = 1 T. The inset in a) shows ln($\Delta \sigma$)vs.$T$ for $H$=0 and a 3D-AL theory fitting extracted from Ref.\cite{rey1}; the inset in b) shows $\delta$$T(H)$ vs. $H$. }
 \label{fig2}
\end{figure}
\begin{figure}[t]
\includegraphics[width=\linewidth]{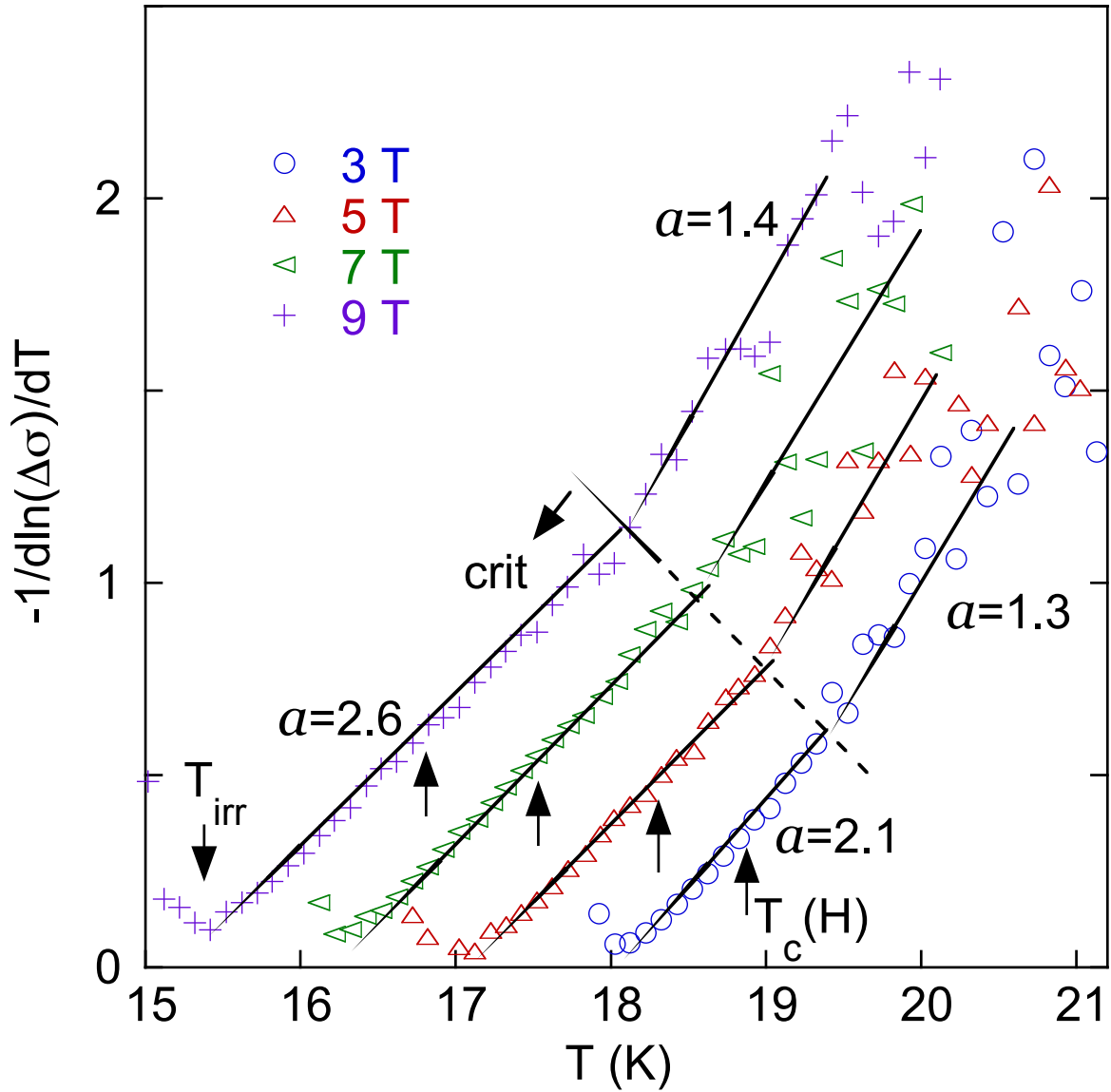}
\caption{-1/dln($\Delta \sigma$)/dT is plotted against T for H$\geq$3T. Vertical arrows up show the position of $T_c(H)$, obtained from the 3D-LLL scaling, in each respective curve.}
 \label{fig3}
\end{figure}

\begin{figure}[t]
\includegraphics[width=\linewidth]{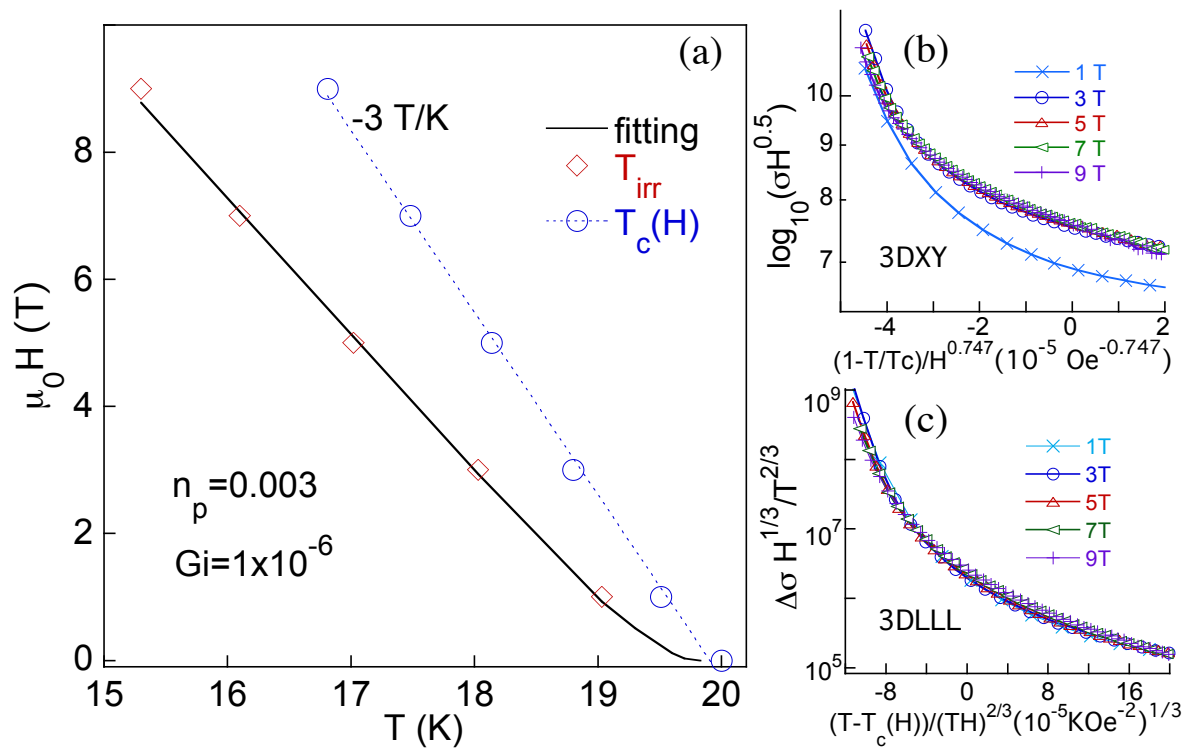}
\caption{a) H vs. T phase diagram for the studied sample. The solid line is a fitting of the irreversibility line to the theory in Ref.\cite{baruch}. b)Results of the 3DXY scaling c) Results of the 3D-LLL scaling. The Y-axis is in units of (Oe$^{1/3}$/ $\Omega$mK$^{2/3}$)}
 \label{fig4}
\end{figure}

Figure 3, shows plots  of -1/dln($\Delta \sigma$)/d$T$ vs $T$  for magnetic fields running from 3 T to 9 T. It is interesting to observe in this figure that all curves exhibit only two different regions, each one with approximately the same exponent (the exponent of the region starting at $T_{irr}$ varies between 2.1 for $H$ = 3 T to 2.6 for $H$ = 9 T, similar to the region starting at $T_{irr}$ for $H$ = 1 T in Fig. 2b, and the exponent of the second region varies between 1.3 and 1.4 similar to the second region above $T_c(H)$ in Fig. 2b), which appears to be a clear effect of the high magnetic field. The absence of a linear region occurring between these two regions, as observed for $H$ = 1 T in Fig. 2b with $a$$\approx$0.88, impeded  us to estimate a value for $T_c(H)$ in each curve of Fig. 3. The second region in the curves of Fig. 3 with $a$$\approx$1.3 are likely to be associated with A-L fluctuations with short-wave length effects (in analogy to the region with $a$$\approx$1.3 in Fig. 2b).\cite{rey1} The first linear regions starting at $T_{irr}$ in the curves of Fig. 3 (and of Fig. 2b) with exponents $a$ varying between 2.1 to 2.6 corresponds to the reversible (flux-flow) regions observed in the resisitivity curves which show a rounding effect (as resisitivity approaches the normal region) due to thermal fluctuations. 

Many types of fluctuations have been proposed to explain the flux-flow region in type II superconductors, including glass-like fluctuations occurring near (above) the glass transition temperature $T_g(H)$ (the glass phase which shows zero resistivity below $T_g$$\approx$$T_{irr}$ may appear from a continuos transition that takes place in the vortex-liquid phase, driven by disorder) \cite{fisher}, low-field 3D-XY fluctuations (driven by phase fluctuations of the complex order parameter as in superfluid $^4$He ) \cite{lobb} and high field LLL fluctuations \cite{ullah,rosenstein}. In the case of LLL fluctuations, the high magnetic field constrict the pairs to the lowest-Landau-Level which acquires an effectively 1D dimension enhancing the importance of fluctuations. As it is shown in Ref.\cite{ullah} the inclusion of the quartic $a \Delta ^4$ term in the G-L Hamiltonian within the LLL-aproximation (this term is only important in the critical region near $T_c(H)$) removes the divergence at the mean field $T_c(H)$ and as a consequence the mean field $H_{c2}(T)$ line becomes a crossover line. 
Since fluctuations of 3DXY type have been used to explain the vortex-liquid phase in resisitivity and magnetization curves in the low field regime \cite{salamon,friesen}, we applied a 3DXY scaling for conductivity to our isofield $\sigma (T)$ curves of Fig. 1, where a plot of  log$_{10}$($\sigma (T)$$H$$^{0.5})$ vs. (1-$T/T_c$)$H$$^{-0.747}$ (where $T_c$ is the only adjustable  parameter, and 1/$2\nu$=-0.747 where $\nu$$_{xy}$=-0.669) are expected to collapse in to a single curve. Figure 4b show this plot with the parameter $T_c$ = 20 K, evidencing that the lower field curve for $H$ = 1 T fail to follow the scaling, which excludes this type of fluctuations.
For the case of glass-like fluctuations occurring near (above) $T_g(H)$ \cite{fisher}, the conductivity is expected to follow $\Delta \sigma \equiv A(T-T_g)^{- a}$ with the critical exponent $a$= $\nu$(2+$z$-$d$) where $\nu$ is the coherence lenght exponent, $z$ is the dynamical critical exponent and $d$ is the dimensionality. This type of fluctuations have been considered in the 3D "gauge glass" model (in the vortex representation) presented in Ref.\cite{wengel} where it is obtained $\nu$=1.3 and $z$$\approx$3.1 producing $a$$\approx$2.7 ($d$=3), which value is close to the values of the exponent $a$ found in Figs. 3 and 2b. This agreement suggests that the first linear regions of Fig. 2b and 3 can be explained in terms of gauge-glass fluctuations \cite{wengel} occurring around (above) $T_{irr}$. In that case, a second order phase transition (possibly a glass phase) should take place at some temperature very close to $T_{irr}$.  

Since the glass transition is driven by disorder, it is important to compare the irreversibility line obtained from $T_{irr}$ values extracted from the curves of Figs. 2b and 3 with the expression developed in Ref.\cite{baruch} for the (glass line) irreversibility line which considers the effect of disorder:
 1-$t$-$b$+2[$n_p$(1-$t$)$^2$$b$/4$\pi$]$^{2/3}$[3/2-(4$\pi$$t$$\sqrt{2Gi}$/($n_p$(1-$t$)$^2$)] = 0 , where $t$=$T$/$T_c$ (we used $T_c$=19.83 K), $b$=$H$/$H_{c2}(0)$ (we used $H_{c2}(0)$ = 42 T), and $n_p$ and $Gi$ (which is related to the Ginzburg number) are fitting parameters defined in Ref\cite{baruch} representing the disorder and the strength of thermal fluctuations respectively (for 1mA, the electric field in the sample is $E$$<$0.01 V/m, allowing us to use the static form of the irreversibility line expression in Ref.\cite{baruch}).  Figure 4 shows a ($H$,$T$) diagram for the irreversibility line where the solid line represents a best fitting to the above expression obtained with $n_p$=0.003 and $Gi$=1x10$^{-6}$ (this value of $Gi$, is two orders of magnitude larger than the value encountered in Ref.\cite{baruch} for NbSe$_2$). Since our value of $n_p$ is larger than the value encountered for NbSe$_2$ in Ref.\cite{baruch} it suggests that disorder is important in the studied sample, supporting the hipothesis of glass-like fluctuations occurring near (above) $T_{irr}$. Despite we did not obtain $M(H)$ magnetization curves to check for the possible existence of a first order melting line and of the peak effect, we mention that $M(H)$ curves obtained in a similar sample of  BaFe$_{1.9}$Ni$_{0.1}$As$_2$ presented in Ref.\cite{SUST} could not resolve any discontinuity in magnetization in the vicinity of $H_{irr}$ and also show that the
second magnetization peak, centered at $H_p$, observed in $M(H)$ curves dissapears as temperature approaches $T_c$ with the corresponding $H_p$ line ending at an isolated point in the (H,T) phase diagram, below the irreversiblity line. It would be interesting to compare values of the exponent $a$ obtained here with those obtained for other superconductors, but unfortunatelly, there is an apparent lack of similar studies in the literature. In any case, the observation of only one possible critical region appears to be in agreement with the predictions of Ref. \cite{benfato} for pnictides.

It is interesting to analyse the behavior of the width, $\delta$$T(H)$, with $H$ of the first linear region (with $a$$>$2) appearing in the curves of Fig. 3 and of Fig. 2b. A plot of the quantity  $\delta$$T(H)$ vs. $H$ is shown in the 
inset of Fig. 2b, and the best fit of the resulting curve produced $\delta$$T$$\propto$$H^{0.652}$ which is in direct agreement with the field-dependent Ginzburg criterium, $\Delta$$T(H)$$\propto$$H^{2/3}$, predicted for three 
dimensional  fluctuations of LLL type.\cite{ullah} To check this possibility, we apply the 3D-LLL scaling law derived in Ref. \cite{ullah} for magneto-conductivity, to the isofield curves of Fig. 1 with $H$ running from 1 to 9 T. Figure 4c shows the results of this scaling applied on data lying in the first linear region of Fig. 3 and of Fig. 2b, where $\Delta \sigma H^{1/3}/T^{2/3}$ is plotted against ($T$-$T_c(H)$)/($TH$)$^{2/3}$ and $T_c(H)$ (representing the 
mean field transition temperature) is an adjustable parameter. The collapse of the curves suggests that LLL-fluctuations also play a role in the first linear regions of the curves of Figs. 3 and 2b. It should be mentioned 
that calculations performed in Ref.\cite{baruch} show that the LLL contribution to conductivity dominates above the glass transition (flux-flow regime) where $\sigma$$_{LLL}$ is a function of $H^{-1/3}T^{2/3}$ and the LLL scaled temperature a function of ($HT$)$^{-2/3}$, which suggests a scaling law for conductivity curves in the vortex-liquid phase similar to the 3D-LLL scaling law used. The LLL scaling procedure allowed to obtain  values of the mean field temperature transition $T_c(H)$ producing a reasonable value for $\mu _0$d$H_{c2}$/d$T$ = -3 T/K. The values of $T_c(H)$ are plotted with values of $T_{irr}$ in Fig. 4a.   
To better visualize the results of this work, we added an arrow in each curve of Fig. 3 and of Fig. 1 indicating the position of $T_c(H)$ obtained from the scaling approach, and a line linking the values of conductivity at the virtual points separating critical and Gaussian regimes in Fig. 3 (dotted line)  and in Fig. 1 (solid line named $T^*$ line). 
\section{Conclusions}
In conclusion, analysis of high-field conductivity curves evidence the existence of only two fluctuation regimes, a possible critical one going from the irreversible temperature to above $T_c(H)$, followed by Aslamazov-Larkin fluctuations in the Gaussian regime. Our analysis suggests that critical fluctuations emerge above $T_{irr}$ under high applied magnetic fields. These critical fluctuations, not resolved for zero magnetic field, appear to account for the excess conductivity above $T_c(H)$, which extends more than 1 K for 9 T. The analysis allowed to estimate with certain accuracy the width of the critical region as well as its associated critical exponent. 
\section*{Acknowledgements}
Work supported by: CNPq and FAPERJ (Brazil); the Spanish MICINN and ERDF (No. FIS2010-19807), and the Xunta de Galicia (Nos. 2010/XA043 and 10TMT206012PR) (Spain); NSFC Program (No. 11004233) and 973 project (No. 2011CBA00110) (P.R. China). We thank discussions with S.L.A. de Queiroz and P. Pureur

\section*{References}.


\end{document}